\documentclass[12pt, epsfig]{article}
\usepackage{graphicx}
  \usepackage{amsthm,amsfonts}
  \usepackage{amsmath}
\newcommand{\bea}   {\begin{eqnarray}}
\newcommand{\eea}   {\end{eqnarray}}

\begin{document}
\renewcommand{\thefootnote}{\fnsymbol{footnote}}

\thispagestyle{empty}

\title{The octonionically-induced ${\cal N}=7$ exceptional \\ $G(3)$ superconformal quantum mechanics}

\author{Francesco
Toppan\thanks{{E-mail: {\em toppan@cbpf.br}}}
\\
\\
}
\maketitle

\centerline{$^{\ast}$
{\it CBPF, Rua Dr. Xavier Sigaud 150, Urca,}}{\centerline {\it\quad
cep 22290-180, Rio de Janeiro (RJ), Brazil.}
~\\
\maketitle
\begin{abstract}
Both the ${\cal N}=7$ superconformal quantum mechanics possessing the exceptional $G(3)$ Lie superalgebra as dynamical symmetry and its associated deformed oscillator with $G(3)$ as spectrum-generating superalgebra are presented.\par
This superconformal quantum mechanics, uniquely defined up to similarity transformations, is obtained via the octonionically-induced ``quasi-nonassociative" method employed to derive the exceptional ${\cal N}=8$ $F(4)$ model. \par
To construct the $G(3)$ theories, the covariant embedding of the $7$-dimensional representation of the Lie algebra $g_2$ within the $8\times 8$ matrices spanning the $Cl(0,7)$ Clifford algebra is derived.\par
The Hilbert space of the $G(3)$ deformed oscillator is given by a $16$-ple of square-integrable functions of a real space coordinate. The spectrum of the theory is computed.
~\\\end{abstract}
\vfill

\rightline{CBPF-NF-005/19
}

\newpage
\section{Introduction}

This work presents the construction of both the ${\cal N}=7$ superconformal quantum mechanics possessing the exceptional $G(3)$ Lie superalgebra as dynamical symmetry and of its associated (via the de Alfaro-Fubini-Furlan construction \cite{dff}) deformed oscillator with $G(3)$ as spectrum-generating superalgebra.
This superconformal quantum mechanics, uniquely defined up to similarity transformations, is obtained via the octonionically-induced ``quasi-nonassociative" method employed \cite{akt} to derive the exceptional ${\cal N}=8$ $F(4)$ superconformal quantum mechanics. Unlike the $F(4)$ theory, the $G(3)$ superconformal mechanics is inherently a quantum theory since the ${\cal N}=7$ worldline  $(1,7,7,1)$ supermultiplet \cite{kuroto} (which carries a $G(3)$ representation \cite{khto}) does not admit a classical Lagrangian defining a world-line sigma-model. 
The Hilbert space of the $G(3)$ deformed oscillator  is given by a $16$-ple of square-integrable functions of the real space coordinate $x$.  The spectrum of the deformed oscillator 
(discrete and bounded from below) is derived.\par
A consequence of the present construction is that both the exceptional Lie superalgebras $G(3)$ and $F(4)$ define their respective, unique, superconformal quantum mechanics. It should be stressed that, as a byproduct, a new  differential matrix representation of $G(3)$ is obtained.\par
It is worth recalling that the $5$ exceptional Lie algebras $g_2$, $f_4$, $e_6$, $e_7$, $e_8$, as well as the $2$ exceptional Lie superalgebras $G(3)$, $F(4)$ in Kac's classification \cite{kac}, are related to the octonions: $g_2$ is the Lie algebra of the group of automorphisms of the octonions, while $f_4$, $e_6$, $e_7$ and $e_8$
are recovered from the octonionic cases in the Freudenthal-Tits magic square construction; the octonionic construction of $G(3)$ and $F(4)$ was presented in \cite{sud}.\par
The ``quasi-nonassociativity" is based on the double role of the octonionic structure constants $C_{ijk}$ which also enter seven $8\times 8$ gamma matrices $\gamma_i$ ($i=1,\ldots,7$) defining the Euclidean $Cl(0,7)$ Clifford algebra. These matrices are induced by the left action of each one of the seven imaginary octonions over a real octonion, see \cite{akt} for details. Since the quasi-nonassociative derivation was presented in \cite{akt}, it is sufficient here  to pinpoint the differences between the $F(4)$ and the $G(3)$ constructions. In both cases $16\times 16$ matrices with differential entries are required. The $R$-symmetry subalgebra of $G_3$ is $g_2$, since the
$G(3)$ even subsector is decomposed as $sl(2)\oplus g_2$. The $14$-generator exceptional Lie algebra $g_2$ admits a $7$-dimensional representation. The key ingredient consists in expressing this representation in terms of 
the antisymmetric, covariant rank-$2$ tensors defined by the $\gamma_i$ matrices. Once this is done (the construction is given in Section {\bf 2}), the $16\times 16$ matrix differential representation of $G(3)$ is carried out by taking into account that the seven supercharges $Q_i$ and their seven superconformal partners ${\widetilde Q}_i$ are labeled by the octonionic vector index $i$.  This construction is presented in Section {\bf 3}.  The deformed oscillator with $G(3)$ spectrum-generating superalgebra, the derivation of its Hilbert space from the $G(3)$ lowest weight reprentations and the computation of its spectrum are all given in Section {\bf 4}. 
A more detailed discussion of the results, together with future outlines, is presented in the Conclusions.

\section{Octonionic covariance and the $g_2$ representation}

The first part of this Section follows \cite{akt}. \par
The octonionic multiplication is defined, for the seven imaginary octonions $e_i$
($i=1,2,\ldots,7$), as
\bea\label{g3oct}
e_i e_j &=& -\delta_{ij}+C_{ijk}e_k
\eea
(here and in the following the sum over repeated indices is understood). The rank-$3$ totally antisymmetric tensor $C_{ijk}$ defines the octonionic structure constants. Two more totally antisymmetric
constant tensors  are compatible with the octonionic multiplication; they are expressed as
$C_{ijkl}$ and $\epsilon_{ijklmnp}$ (their rank is $4$ and $7$, respectively). The following normalizations are assumed
\bea\label{g3constants}
&C_{123}=C_{147}=C_{165}= C_{246}=C_{257}=C_{354}=C_{367}=1,&\nonumber\\
&C_{4567}=C_{2356}=C_{2437}= C_{1357}=C_{1346}=C_{1276}=C_{1245}=1,&\nonumber\\
&\epsilon_{1234567}=1.&
\eea
Due to the relation
\bea
6 C_{ijkl} &=& \epsilon_{ijklmnp}C_{mnp},
\eea
only two of the three constant, totally antisymmetric tensors are independent. \par 
The left action of any given imaginary octonion $e_i$ over a real octonion $x=x_0+x_je_j$ ($x_0, x_j\in {\mathbb R}$)   induces a $8\times 8$ matrix $\gamma_i$ defined by the linear transformation
\bea\label{g3omap}
{\vec x'_{(i)}} &=&\gamma_i {\vec x} \quad \textrm{for} \quad x\mapsto e_ix = x'_{(i)}= -\delta_{ij}x_j+ (x_0\delta_{ik}+C_{ijk}x_j)e_k.
\eea
The $8$ real numbers entering the real octonions $x$, $x'_{(i)}$ are arranged as $8$-component vectors 
${\vec x}$, ${\vec x'_{(i)}}$
so that, e.g., ${\vec x}=(x_0,x_j)^T$.\par
The seven matrices $\gamma_i$ satisfy the $Cl(0,7)$ Euclidean Clifford algebra relations
\bea\label{g3cl07}
\gamma_i\gamma_j+\gamma_j\gamma_i&=& -2\delta_{ij} {\mathbb I}_8,\quad\quad i,j=1,2,\ldots, 7
\eea
(here and in the following ${\mathbb I}_n$ denotes the $n\times n$ identity matrix).\par
The $\gamma_i$ entries are expressed in terms of the octonionic structure constants $C_{ijk}$ through
\bea\label{g3mat}
({\gamma_i})_{LM} &=& \left(\begin{array}{c|c}0&\delta_{im}\\ \hline
-\delta_{il}&C_{ilm}\end{array}\right),
\eea 
where $L$, $M$ take values $L=0,l$ and $M=0,m$, with $l,m=1,2,\dots, 7$. \\
Up to an overall sign, the (\ref{g3mat}) matrices are obtained from the (\ref{g3omap}) map.\par
The totally antisymmetric constants $C_{ijk}$ play a double role. They define the
non-associative octonionic multiplication (\ref{g3oct}) (where, in particular,
$
(e_1e_2)e_4= e_3e_4=-e_5 \neq e_1(e_2e_4)=e_1e_6=e_5
$)
and they enter the matrix representation of the associative $Cl(0,7)$
Clifford algebra.\par
The $Cl(0,7)$ gamma matrices $\gamma_i$ provide a basis to span the $64$-dimensional vector space of $8\times8$ real matrices. The rank $r=0,1,2,3$ products of $r$ different $\gamma_i$ matrices are expressed as 
\bea\label{g3rank7}
&\gamma^{(0)}\equiv {\mathbb I}_8,\quad
\gamma^{(1)}\equiv \gamma_i,\quad
\gamma^{(2)} \equiv  \gamma_i\gamma_j \quad (i<j),\quad
\gamma^{(3)} \equiv \gamma_i\gamma_j\gamma_k\quad (i<j<k).&
\eea
Due to Hodge duality, the product of $7-r$ different matrices $\gamma_i$ is equivalent to the product of matrices of rank $r$. One has that
$\gamma^{(0)}$ and $\gamma^{(3)}$ provide the basis for the $1+35=36$ symmetric $8\times 8$ matrices, while 
$\gamma^{(1)}$ and $\gamma^{(2)}$ provide the basis for the  $7+21=28$ antisymmetric $8\times 8$ matrices.\par
Up to now, this is the construction presented in \cite{akt}.
The extra ingredients required for the construction of the $G(3)$ superconformal quantum mechanics are the following. \par
Scalars (rank-$0$ tensors), vectors (rank-$1$ tensors), rank-$2$ and rank-$3$ tensors are obtained by (partially) saturating the indices labeling the three constant totally antisymmetric tensors (\ref{g3constants}) with the $\gamma_i$ matrices. These tensors provide a basis for the ``octonionic covariant" matrices of given order. It is convenient to illustrate them by presenting the following table
\bea
&\begin{array}{|c|c|c|c|}\hline 
 0A:&C_{ijk}\gamma_i\gamma_j\gamma_k&1&S \\
0B:&C_{ijkl}\gamma_i\gamma_j\gamma_k\gamma_l&1&S\\
0C:&\epsilon_{ijklmnp}\gamma_i\gamma_j\gamma_k\gamma_l\gamma_m\gamma_n\gamma_p&1&S\\ \hline
 1A:&C_{ijk}\gamma_j\gamma_k&7&A\\
1B:&C_{ijkl}\gamma_j\gamma_k\gamma_l&7&S\\
1C:&\epsilon_{ijklmnp}\gamma_j\gamma_k\gamma_l\gamma_m\gamma_n\gamma_p&7&A\\ \hline
 2A:&C_{ijk}\gamma_k&7&A\\
2B:&C_{ijkl}\gamma_k\gamma_l&21&A\\
2C:&\epsilon_{ijklmnp}\gamma_k\gamma_l\gamma_m\gamma_n\gamma_p&21&A\\ \hline
 3A:&C_{ijk}{\mathbb I}_8&1&S \\
3B:&C_{ijkl}\gamma_l&7&A\\
3C:&\epsilon_{ijklmnp}\gamma_l\gamma_m\gamma_n\gamma_p&35&S\\ \hline
\end{array}&
\eea
The third column reports the number of spanning matrices of given type, while their symmetry ($S$) or antisymmetry ($A$) under matrix transposition is specified in the fourth column.\par
Different types of octonionic-covariant matrices do not necessarily determine different matrices. As an example,
the ${\mathbb I}_8$ identity can be expressed either as the $0C$ scalar (${\mathbb I}_8=\frac{1}{7!}\epsilon_{ijklmnp}\gamma_i\gamma_j\gamma_k\gamma_l\gamma_m\gamma_n\gamma_p$) or as the rank-$3$ tensor $3A$ ($C_{ijk}{\mathbb I}_8$). A relevant identification, due to Hodge duality, is
\bea
1C\equiv 2A && (
 \epsilon_{ijklmnp}\gamma_j\gamma_k\gamma_l\gamma_m\gamma_n\gamma_p\sim C_{ijk}\gamma_k).
\eea
There are only two (up to normalization) octonionic-covariant scalar matrices. Besides the identity expressed by ``$0C$", both $0A$ and $0B$ determine the diagonal matrix
\bea
0A\equiv 0B &\sim& diag(7,-1,-1,-1,-1,-1,-1,-1).
\eea
One should further note that the $7$ antisymmetric matrices individuated by $1A$ differ from the $7$ antisymmetric gamma matrices $\gamma_i$, covariantly expressed as $1C$, $2A$ or $3B$.\par
There are three types of octonionically-induced rank-$1$ (vector) matrices, the $7$ symmetric matrices from $1B$, which will be denoted as ``$b_i$'', the $7$ antisymmetric matrices $\gamma_i$ and the $7$ antisymmetric matrices from $1A$, which will be denoted as ``$m_i$". One can set
\bea\label{bivectors}
b_i &=& \frac{1}{24}C_{ijkl}\gamma_j\gamma_k\gamma_l, \quad\quad {\textrm{so that}}\quad (b_i)_{LM}= \left(\begin{array}{cc}0&\delta_{im}\\ 
\delta_{il}&0\end{array}\right),\nonumber\\
m_i&=&\frac{1}{2}C_{ijk}\gamma_j\gamma_k.
\eea
A more convenient basis to express the generic rank-$1$ antisymmetric matrices $a_i$ is to present them as linear combinations of $\gamma_i$, $n_i$ ($a_i=w_1\gamma_i+w_2n_i$, with $w_1,w_2\in {\mathbb R}$),  where $n_i$
is introduced as
\bea\label{nivectors}
n_i &=& \frac{1}{4}(m_i+3\gamma_i), \quad \quad{\textrm{so that}}\quad (n_i)_{LM}= \left(\begin{array}{cc}0&0
\\0&C_{ilm}\end{array}\right).
\eea
The matrices $n_i$ are nonvanishing in the $7\times 7$ lower-right block only.\par
The most general octonionically-induced rank-$2$ antisymmetric matrices $a_{ij}$ are given by a linear combination of
$2A$, $2B$ and $2C$. One can set
\bea
a_{ij}& = & z_1 C_{ijk}\gamma_k+z_2C_{ijkl}\gamma_k\gamma_l+z_3\epsilon_{ijklmnp}\gamma_k\gamma_l\gamma_m\gamma_n\gamma_p, \quad {\textrm{with}}\quad z_1,z_2,z_3\in {\mathbb R}.
\eea
One restricts the coefficients $z_1,z_2,z_3$ by imposing two conditions. The first one is the request that
the $a_{ij}$ matrices are nonvanishing only in the $7\times 7$ lower-right block.  This condition is fulfilled
if $z_3$ is constrained to satisfy
\bea\label{7times7constraint}
z_3&=& -\frac{1}{120}z_{1}-\frac{1}{30}z_2.
\eea
This condition leaves at most $21$ linearly independent $a_{ij}$ matrices.\par
The second condition comes from satisfying the covariant constraint 
\bea\label{g2constraint}
C_{ijk}a_{jk}&=&0.
\eea
This condition is satisfied if $z_1$ is set to vanish:
\bea
z_1&=& 0.
\eea
The (\ref{g2constraint}) constraint implies $7$ relations, leaving $21-7=14$ linearly independent matrices $a_{ij}$. These matrices realize, in their $7\times 7$ lower-right block, the fundamental $7$-dimensional representation of the $14$ generators of the exceptional $g_2$ Lie algebra. By suitably setting the overall normalization, one can express the $14$ linearly independent matrices as $r_{ij}$, defined as
\bea\label{rij}
r_{ij}&=&\frac{1}{2}C_{ijkl}\gamma_k\gamma_l-\frac{1}{60}\epsilon_{ijklmnp}\gamma_k\gamma_l\gamma_m\gamma_n\gamma_p.
\eea
The $7$ rank-$1$ matrices $n_i$ introduced in (\ref{nivectors}) can be expressed in the basis of the $a_{qj}$ matrices satisfying
the (\ref{7times7constraint}) condition. One gets
\bea
n_i&=& \frac{1}{6}C_{iqj}\cdot(\frac{3}{4} C_{qjk}\gamma_k-\frac{1}{8}C_{qjkl}\gamma_k\gamma_l-\frac{1}{480}\epsilon_{qjklmnp}\gamma_k\gamma_l\gamma_m\gamma_n\gamma_p).
\eea 
The set of $r_{ij}$ and $n_i$ matrices produces the $21$ linearly independent antisymmetric  matrices with
nonvanishing $7\times 7$ lower-right block (and vanishing otherwise). Schematically, their commutation relations satisfy
\bea\label{schematic}
&[r,r]\sim r, \qquad [r,n]\sim n,\qquad [n,n]\sim r+n.&
\eea
The matrices $n_i$ belong to the $7$-dimensional representation of the $g_2$ algebra. The whole set of matrices
$r_{ij}$, $n_i$ realize the $7$-dimensional matrix representation of the $so(7)$ Lie algebra (while the commutators $[\gamma_i,\gamma_j]$ realize the $8$-dimensional matrix representation of $so(7)$ since their first columns/rows are nonvanishing).\par
The following remark is worth to be stressed: the octonionically-induced covariant decomposition of matrices produces a nice embedding of the $7$-dimensional matrix representations of both $g_2$ and $so(7)$ inside the $8$-dimensional representation of the $Cl(0,7)$ Clifford algebra.\par
The explicit commutators among the $r_{ij}$, $n_i$ matrices, schematically presented in (\ref{schematic}), are covariantly written as
\bea\label{octcov}
\relax [r_{ij}, r_{kl}]&=& a(\delta_{ik}r_{jl}-\delta_{il}r_{jk}-\delta_{jk}r_{il}+ \delta_{jl}r_{ik})+\nonumber\\
&&b(\delta_{ik} C_{jlmn}-\delta_{il}C_{jkmn}-\delta_{jk}C_{ilmn}+\delta_{jl}C_{ikmn})r_{mn}+\nonumber\\
&& (3-a-\frac{b}{2})(C_{ijkm}r_{ml}-C_{ijlm}r_{mk}-C_{klim}r_{mj}+C_{kljm}r_{mi})+\nonumber\\
&&(4-a-2b) C_{ijm}C_{kln}r_{mn},\nonumber\\
\relax [r_{ij},n_k]&=&4\delta_{ik}n_j-4\delta_{jk}n_i+2C_{ijkl}n_l,\nonumber \\
\relax [n_i,n_j]&=& \frac{1}{2}r_{ij}+C_{ijk}n_k.
\eea
As a consequence of the $C_{ijk}r_{jk}=0$ constraint, on the right hand side of the first equation two real parameters $a,b$ can be arbitrarily chosen since any pair of their selected values define the same generator. The first equation gives the octonionic-covariant expression of the structure constants of the $g_2$ Lie algebra. The whole set of three equations gives the structure constants of $so(7)$.\par
The second equation in (\ref{octcov}) shows that the $n_i$'s belong to a representation of $g_2$.
The covariant rank-$1$ matrices $\gamma_i$, $n_i$, $b_i$ satisfy the same $7$-dimensional representation of $g_2$. Indeed, in all these cases  one gets
\bea
[r_{ij},v_k]&=&4\delta_{ik}v_j-4\delta_{jk}v_i+2C_{ijkl}v_l,
\eea
where the $v_i$'s can be respectively replaced by  $\gamma_i$, $n_i$ or $b_i$.

\section{The $G(3)$ superconformal quantum mechanics}

The exceptional, finite, $G(3)$ Lie superalgebra can be interpreted, see \cite{khto}, as a one-dimensional superconformal algebra with ${\cal N}=7$ extension. $G(3)$ admits a $5$-grading decomposition, given by
\bea\label{dec}
G(3) &=& {\cal G}_{-1}\oplus{\cal G}_{-\frac{1}{2}}\oplus {\cal G}_0\oplus{\cal G}_{\frac{1}{2}}\oplus {\cal G}_1.
\eea
The half-integer sectors ${\cal G}_{\pm\frac{1}{2}}$ are odd; the integer sectors ${\cal G}_0$, ${\cal G}_{\pm 1}$ are even. \\
The (anti)commutators (collectively denoted as ``$[\cdot,\cdot\}$" when needed) respect the above decomposition, so that
\bea
[{\cal G}_{s_1},{\cal G}_{s_2}\}&\subset&{\cal G}_{s_1+s_2}, \qquad {\textrm {for}} \quad s_1,s_2=0,\pm{\frac{1}{2}}, \pm 1.
\eea
Each ${\cal G}_{\pm 1}$ sector is spanned by a single generator, respectively denoted as ``$H,K$", with $H\in {\cal G}_1$ and $K\in {\cal G}_{-1}$. The ${\cal G}_0$ sector is a subalgebra, given by the direct sum
\bea
{\cal G}_0&=& u(1)\oplus g_2.
\eea
In application to superconformal mechanics, the exceptional Lie algebra $g_2$ is known as the ``$R$-symmetry". The $u(1)$ generator is denoted as ``$D$". It corresponds to the dilatation operator. The set of $D,H,K$ generators close an $sl(2)$ subalgebra, with  $D$ as the Cartan element and $H$ ($K$) as the positive (negative) root. \par
The $5$-grading decomposition (\ref{dec}) is related to the scaling dimension of the generators, defined by their commutators with $D$. Indeed, one gets
\bea
[D, Z_s] = is Z_s, &&  \forall Z_s\in {\cal G}_s.
\eea
The odd sector ${\cal G}_{\frac{1}{2}}$ is spanned by seven supercharges, denoted as ``${Q_i}$" ($i=1,2, \ldots, 7$), while the  ${\cal G}_{-\frac{1}{2}}$ sector is spanned by their $7$ conformal superpartners, denoted as ``${\widetilde Q}_i$". \par
The positive sector ${\cal G}_{>0}={\cal G}_{\frac{1}{2}}\oplus {\cal G}_1$ is isomorphic to the ${\cal N}=7$-extended worldline superalgebra (see \cite{wit}), defined by the (anti)commutators
\bea\label{worldsusy}
\{Q_i,Q_j\} = 2\delta_{ij} H, && [H, Q_i]=0.
\eea
The nonvanishing (anti)commutators of $G(3)$ are presented for completeness;  due to the octonionic covariance the structure constants are expressed, as in formula (\ref{octcov}),  in terms of the constant octonionic tensors. In this presentation the $g_2$ subalgebra generators are given in terms of the $R_{ij}$ antisymmetric tensor satisfying the constraints
\bea
C_{ijk}R_{jk}&=&0.
\eea
The nonvanishing (anti)commutators are
\bea\label{g3algebra}
&&[D,H]=iH,\qquad\quad ~[D,K]=-iK,\qquad ~[D, Q_i]=\frac{i}{2}Q_i,\qquad [D,{\widetilde Q}_i]=-\frac{i}{2}{\widetilde Q}_i,\nonumber\\
&&[H,K]=-2iD, \qquad [H, {\widetilde Q}_i]= -iQ_i, \qquad [K, {Q}_i]= i{\widetilde Q}_i,\nonumber\\
&&[R_{ij}, R_{kl}] = \frac{3i}{4}(C_{ijkm}r_{ml}-C_{ijlm}R_{mk}-C_{klim}R_{mj}+C_{kljm}R_{mi})+i C_{ijm}C_{kln}R_{mn},\nonumber\\
&&[R_{ij}, Q_k] = i(\delta_{ik}Q_j-\delta_{jk}Q_i)+\frac{i}{2}C_{ijkl}Q_l, \qquad [R_{ij}, {\widetilde Q}_k ] = i(\delta_{ik}{\widetilde Q}_j-\delta_{jk}{\widetilde Q}_i)+\frac{i}{2}C_{ijkl}{\widetilde Q}_l,
\nonumber\\
&&\{Q_i, Q_j\} = 2\delta_{ij}H,\qquad \{{\widetilde Q}_i, {\widetilde Q}_j\}= 2\delta_{ij}K, \qquad \{ Q_i, {\widetilde Q}_j\}=2D\delta_{ij}+R_{ij}.
\eea
The structure constants of the $g_2$ subalgebra are here presented by setting equal to zero the parameters $a,b$ entering the right hand side of the first equation in (\ref{octcov}): $a=b=0$. Please note that the $g_2$ generators are now given in capitalized form ($R_{ij}$ instead of $r_{ij}$).

\subsection{The differential representation}

The construction of the superconformal quantum mechanics with $G(3)$ as dynamical symmetry requires the following steps. One introduces the space coordinate $x$ and its associated derivative $\partial_x$ with assigned scaling dimensions
\bea
[x] = -\frac{1}{2}, && [\partial_x]=\frac{1}{2}.
\eea
The generators entering (\ref{g3algebra}) are realized by Hermitian differential operators (since no confusion arises, the same symbol will be used to denote a $G(3)$ generator and its differential representative).    As customary, the fermionic operators $Q_i$, ${\widetilde Q}_i$ are block-antidiagonal. For this reason  the size of the matrices 
introduced in Section {\bf 2} has to be doubled. Therefore, the differential representation is given by $16\times 16$ matrices with differential entries. The superconformal Hamiltonian is given by $H\in{\cal G}_1$.  Its $s=1$ scaling
property implies that it is a diagonal operator with an inverse square potential. With a suitable normalization of its Laplacian term, $H$ is of the form
\bea\label{confham}
H&=& -\frac{1}{2}{\partial_x}^2\cdot {\mathbb I}_{16}+ \frac{1}{x^2}V,\qquad\quad  V=diag(V_0,V_1,\ldots, V_{15}),
\eea
where $V$ is a diagonal matrix whose $V_I$ ($I=0,\ldots, 15$) real entries have to be determined.  \par
The conformal partner of the Hamitonian is the operator $K\in {\cal G}_{-1}$. It is expressed by the quadratic term
\bea\label{kop}
K&=& \frac{1}{2}x^2\cdot{\mathbb I}_{16}.
\eea
The dilatation operator $D$, obtained from the $[H,K]$ commutator,  is
\bea\label{dop}
D&=&-\frac{i}{2}(x\partial_x+\frac{1}{2})\cdot{\mathbb I}_{16}.
\eea
The remaining $16\times 16$ matrix differential operators are constructed in terms of the $8\times 8$ matrices $\gamma_i$, $b_i$, $n_i$, $r_{ij}$ introduced in (\ref{g3mat}), (\ref{bivectors}), (\ref{nivectors}) and (\ref{rij}), respectively. The following $2\times 2$ matrices are further introduced to produce $16\times 16$ matrices as tensor products. It is convenient to set
{\footnotesize}{\bea
&{\mathbb I}_2=\left(\begin{array}{cc} 1&0\\0&1\end{array}\right),\quad X=\left(\begin{array}{cc} 1&0\\0&-1\end{array}\right),\quad Y=\left(\begin{array}{cc} 0&1\\1&0\end{array}\right),\quad A=\left(\begin{array}{cc} 0&1\\-1&0\end{array}\right).&
\eea
}}
In octonionic-covariant notation, the most general $16$-dimensional representations of $g_2$ can be expressed, in convenient normalization, either as the Hermitian matrices
\bea\label{g2subalgebra}
R_{ij} &=& \frac{i}{4} {\mathbb I}_2\otimes r_{ij},
\eea
or as the Hermitian matrices
\bea\label{g2pmsubalgebras}
R^\pm_{ij} &=& \frac{i}{8}({\mathbb I}_2\pm X)\otimes r_{ij}
\eea
(in the latter case $R^+_{ij}\leftrightarrow R^-_{ij}$ are mutually recovered via a similarity transformation). \\ The differential representation of $G(3)$ forces the $g_2$ subalgebra to be represented by (\ref{g2subalgebra}), while the (\ref{g2pmsubalgebras}) solutions have to be discarded because they do not allow the $Q_i$, ${\widetilde Q}_i$ operators introduced below to be in a representation of $g_2$.\par
The most general block-antidiagonal, Hermitian operators of scaling dimension $s=\frac{1}{2}$ ($s=-\frac{1}{2}$) and carrying a vector index $i$ are expressed as $Q_i$ (${\widetilde Q}_i$); they are given by
\bea\label{qiqtildei}
Q_i=-\frac{i}{\sqrt{2}}\left(A_i\partial_x+\frac{B_i}{x}\right), &\qquad&
{\widetilde Q}_i = \frac{x}{\sqrt{2}}C_i,
\eea
where the matrices $A_i$, $B_i$ are respectively given by the linear combinations 
\bea
A_i&=& k_1 A\otimes \gamma_i +k_2A\otimes n_i+k_3Y\otimes b_i+N_F\cdot (k_4 Y\otimes \gamma_i+k_5Y\otimes n_i+k_6A\otimes b_i),\nonumber\\
B_i&=&  {\widetilde k}_1 Y\otimes \gamma_i +{\widetilde k}_2Y\otimes n_i+{\widetilde k}_3A\otimes b_i+N_F\cdot ({\widetilde k}_4 A\otimes \gamma_i+{\widetilde k}_5A\otimes n_i+{\widetilde k}_6Y\otimes b_i).\nonumber\\
&&
\eea
The matrix $N_F$ is the fermion parity operator. It is defined as
\bea\label{fermionparity}
N_F&=& X\otimes {\mathbb I}_8.
\eea
The real coefficients $k_1,\ldots , k_6$ and ${\widetilde k}_1, \ldots , {\widetilde k}_6$ have to be determined
by requiring the closure of the (\ref{g3algebra}) (anti)-commutators.\par
The matrices $C_i$ entering the second equation of (\ref{qiqtildei}) are expressed by the same linear combinations
as the matrices $A_i$. Without loss of generality the requirement from (\ref{g3algebra}) that, at given $i$, the anticommutator 
$\{Q_i,{\widetilde Q}_i\}$ would be proportional to the dilatation operator $D$, implies that one can set
\bea
C_i&=&A_i.
\eea
The matrices $A_i$, $B_i$ must fulfill the following conditions
\bea\label{conditiona}
\{A_i, A_j\}&=& 2\delta_{ij}{\mathbb I}_{16},\nonumber\\
\{A_i, B_j\} +\{A_j, B_i\}&=& 0,\nonumber\\
\{B_i, B_j\}-A_iB_j-A_jB_i&=&0,
\eea
resulting from the closure of the worldline superalgebra (\ref{worldsusy}) with the identification of the Hamiltonian $H$ given in (\ref{confham}). This identification further implies that
\bea\label{conditionb}
V&=& -\frac{1}{2}(B_i^2-A_iB_i) \qquad {\textrm{for any given }}\quad i=1,2,\ldots, 7.
\eea \par The anticommutators $\{Q_i, {\widetilde Q}_j\}$, for $i\neq j$, should produce antisymmetric matrices proportional to the $g_2$ $R$-symmetry generators $R_{ij}$ from (\ref{g2subalgebra}), so that
\bea\label{conditionc}
A_iA_j+B_iA_j+A_jB_i&\propto&R_{ij}.
\eea
Solving the constraints (\ref{conditiona},\ref{conditionb},\ref{conditionc}) guarantees the closure of the $G(3)$ superconformal algebra. The solutions are obtained with the following steps.\par
The first equation in (\ref{conditiona}) implies that $k_4, k_5, k_6$ have to be set to
\bea
&k_4= \frac{1}{2}(\varepsilon_1-\varepsilon_2-2k_1), \quad k_5=\frac{1}{2}(-\varepsilon_1+\varepsilon_2+2\varepsilon_3-2k_2),\quad k_6=\frac{1}{2}(-2k_3+\varepsilon_1+\varepsilon_2),&\nonumber\\&&
\eea
where $\varepsilon_a$ are three independent signs ($\varepsilon_a=\pm 1$ for $a=1,2,3$), while $k_1,k_2,k_3$ remain arbitrary real numbers.\par
The second equation in (\ref{conditiona}) is automatically satisfied, while relations for ${\widetilde k}_i$'s are obtained from the third equation in (\ref{conditiona}) and from (\ref{conditionb},\ref{conditionc}). The linearity of the
(\ref{conditionc}) constraint makes more convenient to solve it first. Then, the two other conditions unambiguously fix all remaining ${\widetilde k}_i$'s. \par
This analysis has to be repeated for each one of the $8$ different cases corresponding to the $3$ sign assignments
of the $\varepsilon_a$'s. It is easily checked that in all these cases the same diagonal matrix $V$, defining the potential term of the (\ref{confham}) Hamiltonian, is recovered. This means that $V$ does not depend on the 
arbitrary choices of  the real parameters $k_1, k_2,k_3$ and of the signs $\varepsilon_1,\varepsilon_2,\varepsilon_3$. Therefore, the superconformal Hamiltonian $H$ is uniquely determined. It is given by
\bea\label{hamop}
H= -\frac{1}{2}{\partial_x}^2\cdot {\mathbb I}_{16}+ \frac{1}{x^2}V,&\quad& V=diag(V_0,V_1,\ldots, V_{15}),\quad{\textrm{with}}
\nonumber\\ && V_0=V_8=1, \quad V_i=V_{8+i}=0\quad{\textrm{for}} \quad i=1,\ldots, 7.
\eea
The most suitable presentation of the $G(3)$ differential matrix representation is obtained by setting
\bea
k_1=1,\qquad k_2=k_3=k_4=k_5=k_6=0&& ({\textrm{therefore}} \quad \varepsilon_1=-\varepsilon_2=\varepsilon_3=1).  
\eea
With this choice of parameters one gets
\bea\label{qiop}
Q_i&=&-\frac{i}{\sqrt{2}}\left(A\otimes \gamma_i\partial_x-\frac{A\otimes b_i}{x}\right)
\eea
and
\bea\label{qtildeiop}
{\widetilde Q}_i&=&\frac{x}{\sqrt 2}A\otimes \gamma_i.
\eea
The operators $H, K, D, R_{ij}, Q_{i}, {\widetilde Q}_i$, respectively introduced in (\ref{hamop},\ref{kop},\ref{dop},\ref{g2subalgebra},\ref{qiop},\ref{qtildeiop}), close the $G(3)$ superalgebra (anti)commutators  (\ref{g3algebra}). \par
The coupling constants of the square inverse potential are expressed by the entries of the diagonal matrix $V$ presented in (\ref{hamop}). In the octonionic-covariant formalism, $V$ is given by the scalar combination
\bea\label{Vpot}
V&=& {\frac{1}{8}}{\mathbb I}_{16}-\frac{1}{48} C_{ijk}\Gamma_i\Gamma_j\Gamma_k \Gamma_8\Gamma_9, \quad
{\textrm{with}}\quad \Gamma_i= A\otimes\gamma_i, \quad \Gamma_8=Y\otimes {\mathbb I}_8,\quad \Gamma_9= X\otimes {\mathbb I}_8.\nonumber\\
&&
\eea
The supertrace of $V$ is vanishing:
\bea
str (V) &=& 0.
\eea

\section{The $G(3)$ deformed oscillator}

Following the de Alfaro-Fubini-Furlan \cite{dff} construction, the deformed matrix oscillator with $G(3)$ as spectrum-generating superalgebra is given by the Hamiltonian $H_{osc}$,
\bea
H_{osc} &=& H+K,
\eea
with $H$, $K$ respectively introduced in (\ref{hamop}), (\ref{kop}). Therefore
\bea
&H_{osc}= -\frac{1}{2}{\partial_x}^2\cdot {\mathbb I}_{16}+ \frac{1}{x^2}V +\frac{1}{2}x^2\cdot {\mathbb I}_{16},&\nonumber\\&{\textrm{with}} \quad V=diag(V_0,V_1,\ldots, V_{15}),\quad V_0=V_8=1, \quad V_i=V_{8+i}=0\quad{\textrm{for}} \quad i=1,\ldots, 7.&\nonumber\\&&
\eea
This matrix Hamiltonian corresponds to the direct sum of $14$ undeformed oscillators plus $2$ oscillators which are deformed by the presence of the extra $\frac{1}{x^2}$ potential term.\par
$7$ pairs of deformed creation ($a_i^+$) and annihilation ($a_i^-$) operators are introduced through the positions
\bea
a_i^\pm &=&{\widetilde Q}_i\mp i Q_i=\frac{1}{\sqrt 2}\left( A\otimes \gamma_i(\mp\partial_x+x)\pm\frac{A\otimes b_i}{x}\right) .
\eea
They satisfy the commutation relations
\bea
[H_{osc}, a_i^\pm] &=& \pm a_i^{\pm}.
\eea
For any given $i$ the $H_{osc}$ Hamiltonian is expressed by the anticommutators (no summation over the repeated indices)
\bea
H_{osc} &=& \frac{1}{2}\{ a_i^+, a_i^-\}.
\eea
For any given $i$ the commutator of the creation/annihilation operators produces a deformed (due to the presence on the right hand side of a Klein operator) Heisenberg algebra:
\bea
[a_i^+, a_i^-] &=& {\mathbb I}_{16} + 2S_i,
\eea
where $S_i$ is a diagonal matrix.
For any $i=1,\ldots, 7$, the matrix $S_i$ is a Klein operator since it satisfies
\bea
{S_i}^2={\mathbb I}_{16}, &\quad& \{S_i, a_i^+\}=\{S_i, a_i^-\}=0.
\eea
Explicitly, one has
\bea
S_i&=& diag(s_0,\ldots, s_{15}), \quad {\textrm{with}} \quad s_0=s_8=-1,\quad s_j=s_{8+j}=\delta_{ij}\quad{\textrm{for}} \quad j=1,\ldots, 7.\nonumber\\
&&
\eea
Each one of the seven annihilation operators $a_i^-$ defines $16$ lowest weight vectors $\Psi_{lwv}$ as solutions of the
equation
\bea\label{lwv}
a_i^-\Psi_{lwv}&=&0.
\eea
The sixteen solutions of (\ref{lwv}) are denoted as $\Psi_i^{(J)}$ ($J=0,1,\ldots, 15$); only the $J$-th component of
the $\Psi_i^{(J)}$ vector is nonvanishing. The lowest weight vectors 
are given by
\bea
{\Psi_i^{(J)}} &=& (\Psi_{i,0}^{(J)},\Psi_{i,1}^{(J)},\ldots, \Psi_{i, 15}^{(J)})^T, \quad{\textrm{with}}\quad\Psi_{i,K}^{(J)}=\delta_{JK}\psi_{i,K}
\eea 
and, up to a proportionality factor, 
\bea\label{lwvexplicit}
&\psi_{i,0}=\psi_{i,8}=\frac{1}{x}e^{-\frac{1}{2}x^2},\quad\psi_{i,i}=\psi_{i,8+i}=xe^{-\frac{1}{2}x^2}, \quad \psi_{i,K}=e^{-\frac{1}{2}x^2}\quad (K\neq 0,8,i,8+i).&\nonumber\\&&
\eea 
It follows in particular that, irrespective of the value of $i=1,\ldots, 7$,  the vectors $\Psi_i^{(0)}$ denote the same wave function
(similarly, independently of $i$,  the vectors $\Psi_i^{(8)}$ denote another unique wave function).\par
Each lowest weight vector $\Psi_i^{(J)}$ defines a lowest weight representation spanned by the vectors
$(a_1^+)^{n_1} (a_2^+)^{n_2} \ldots (a_7^+)^{n_7}\Psi_i^{(J)}$, with $n_1,\ldots, n_7$ arbitrary non-negative integers.
The Hilbert space of the model is given by a direct sum of those lowest weight representations which allow normalized wave functions. The selection of the admissible lowest representations proceeds as follows.
\par
At first one has to notice that $\psi_{i,0}$, $\psi_{i,8}$ given in (\ref{lwvexplicit}) do not produce normalized square-integrable functions due to the singular integration $\sim \int dx \frac{1}{x^2}$ at the origin. Therefore, the lowest weight representations induced by the lowest weight vectors $\Psi_i^{(0)}$ and $\Psi_i^{(8)}$ are not admissible. A further analysis similarly proves that all lowest weight representations induced by the gaussian lowest weight vectors
$\psi_{i,K}$ with $K\neq 0,8,i, 8+i$, are also not admissible. Indeed, each such a gaussian wave function produces
$\Psi_i^{(0)}$, $\Psi_i^{(8)}$ as descendant states via the application of some creation operator $a_j^+$. Let's take, as an example, $\Psi_2^{(1)}$, leading to the gaussian function $\psi_{2,1}$ for $i=2$, $K=J=1$. A straightforward check shows that 
\bea
a_1^+\Psi_2^{(1)}&\propto& \Psi_i^{(8)} \quad \forall i=1,\ldots, 7.
\eea
A similar construction applies to all gaussian functions entering (\ref{lwvexplicit}). Their lowest weight representations under the action of the $7$ creation operators $a_j^+$ produce non normalizable wave functions
and, therefore, cannot be used to construct a Hilbert space.\par
It turns out that the Hilbert space can be defined to be the direct sum of the $14$ lowest weight vectors $\Psi_i^{(i)}$, $\Psi_i^{(8+i)}$ whose component wave functions are the odd-parity (under $x\leftrightarrow -x$ transformation) functions $xe^{-\frac{1}{2}x^2}$ entering (\ref{lwvexplicit}).\par
These lowest weight vectors are energy eigenstates with common (degenerate) energy eigenvalue $E=\frac{3}{2}$.\par
The diagonal fermion parity operator $N_F$ introduced in (\ref{fermionparity}) defines bosonic (fermionic) states as its $+1$
($-1$) eigenspaces. Since $N_F$ commutes with the Hamiltonian $H_{osc}$,
\bea
[N_F,H_{osc}]&=&0,
\eea
it can be used to introduce a superselected Hilbert space ${\cal H}$. The normalizable vectors $\Psi\in{\cal H}$ satisfy
the superselection condition
\bea
P\Psi&=& \Psi,
\eea
where $P$ is the projection operator ($P^2=P$) given by
\bea
P&=& N_F\cdot e^{\pi i (H_{osc}-\frac{3}{2})}.
\eea
The superselected Hilbert space admits a bosonic, $7$ times degenerate, vacuum defined by the states
$\Psi_i^{(i)}$ with $i=1,\ldots,7$.  It is convenient, for simplicity,  to denote the vacuum states as $|\Psi_i\rangle=\Psi_i^{(i)}$.\par
 The energy spectrum of the theory is given by the eigenvalues
\bea
E_n &=& \frac{3}{2}+n,\qquad n \in {\mathbb N}_0,
\eea
The excited states ($n>0$) are obtained by applying the creation operators $a_j^+$ which, by construction, anticommute with the fermion parity operator,
\bea
\{N_F, a_j^+\}&=&0.
\eea
Accordingly, the eigenstates of energy level $E_n$ are bosonic (fermionic) if $n$ is even (odd).\par
The $7$ creation operators $a_j^+$ satisfy the ``soft supersymmetry version", see \cite{cht}, of the ${\cal N}=7$
worldline superalgebra (\ref{worldsusy}):
\bea\label{soft}
\{a_i^+,a_j^+\}= 2\delta_{ij}Z, &\quad& [Z,a_i^+]=0,\qquad i,j=1,\ldots, 7,
\eea
with $Z$ given by
\bea
Z&=&-H+K-2iD,
\eea
for $H,K,D$ respectively introduced in (\ref{hamop},\ref{kop},\ref{dop}).
The notion of soft supersymmetry refers to the fact that the operator $Z$ is not a Hamiltonian, but a raising operator.\par 
The determination of the degeneracy of the spectrum of the excited states proceeds similarly as for the $F(4)$
\cite{akt} oscillator. At the first ($n=1$) excited level, $7\times 7=49$ excited states $a_i^+|\Psi_j\rangle$ can be written. The $7$ cases corresponding to $i=j$ produce the same eigenstate. The $42$ remaining  cases with $i\neq j$ determine  a total number of $7$ inequivalent eigenstates (each one obtained from $6$ different combinations; as an example, up to normalization, the same eigenstate is expressed as $a_1^+|\Psi_2\rangle$,  $a_2^+|\Psi_1\rangle$,
$a_4^+|\Psi_5\rangle$, $a_5^+|\Psi_4\rangle$, $a_6^+|\Psi_7\rangle$ or $a_7^+|\Psi_6\rangle$). This leaves a total number of $8$ fermionic eigenstates of energy $E=\frac{5}{2}$.\par
The construction gets repeated at each excited level producing $8$ degenerate states at any given integer $n>0$.
The semi-infinite $(7;8;8;8;\ldots)$ tower of energy eigenstates is a consequence of the ${\cal N}=7$ $(7,8,1)$ worldline supermultiplet \cite{kuroto} applied to the soft superalgebra (\ref{soft}).\par
Let's summarize these results: the vacuum energy $E_{vac}$, the excited energy eigenstates $E_n$ and their respective  degeneracies $d(E_{vac})$, $d(E_n)$ are
\bea
&E_{vac}=\frac{3}{2}, \quad d(E_{vac})= 7,\quad E_n = E_{vac}+n, \quad d(E_n)=8,\quad n=1,2,\ldots .&
\eea
The $7$ degenerate, bosonic, normalized ground energy wave functions $|\Psi_i\rangle$ are
\bea
|\Psi_i\rangle&=& ({\overline \psi}_{i,0},{\overline \psi}_{i,1},\ldots, {\overline \psi}_{i,15})^T,
\eea
where
\bea
{\overline \psi}_{i,K} &=&\delta_{iK}{\frac{2^{\frac{1}{2}}}{{\pi^{\frac{1}{4}}}}}xe^{-\frac{1}{2}x^2},\quad {\textrm{for}}\quad i=1,\ldots, 7, \quad K=0,1,\ldots, 15.
\eea
A convenient (unnormalized) presentation for the $8$ distinct fermionic wave functions $\Psi_I^{[1]}$, where $I=0,1,\ldots, 7$, of the first ($n=1$, $E=\frac{5}{2}$) excited level
is given by
\bea
&\Psi_0^{[1]}=a_1^+|\Psi_1\rangle,~\Psi_1^{[1]}=a_2^+|\Psi_3\rangle,~\Psi_2^{[1]}=a_3^+|\Psi_1\rangle,~
\Psi_3^{[1]}=a_1^+|\Psi_2\rangle,\nonumber\\
& \Psi_4^{[1]}=a_7^+|\Psi_1\rangle,~\Psi_5^{[1]}=a_1^+|\Psi_6\rangle,~
\Psi_6^{[1]}=a_5^+|\Psi_1\rangle,~\Psi_7^{[1]}=a_1^+|\Psi_4\rangle.&
\eea
The unique nonvanishing component wave function of $\Psi_0^{[1]}$ is at the $K=8$ position and proportional to $x^2e^{-\frac{1}{2}x^2}$, while the unique nonvanishing component wavefunction of $\Psi_i^{[1]}$ is at the
$K=8+i$ position and proportional to $(x^2-1)e^{-\frac{1}{2}x^2}$.\par
The (unnormalized) $8$ distinct wave-functions  $\Psi_I^{[n+1]}$ of the $(n+1)$-th excited level, for $n\geq 1$, can be expressed via the recursive
formula
\bea
\Psi_0^{[n+1]}=a_1^+\Psi_1^{[n]}, &\quad& \Psi_i^{[n+1]}=a_i^+\Psi_0^{[n]} \quad {\textrm{for}} \quad i=1,2,\ldots, 7.
\eea
  
It is worth to compare the $G(3)$ energy spectrum  with the one obtained from the exceptional ${\cal N}=8$ $F(4)$ deformed oscillator. The ${\cal N}=7$ $G(3)$ case corresponds to a shifted version with the same degeneracy of the vacuum
energy and of the excited states, but with different vacuum energy. The vacuum energy of the $F(4)$ model is
$E_{vac}^{F(4)}=\frac{2}{3}$. \par
Another difference with respect to the $F(4)$ case concerns the component wave functions. Some of the component wave functions
of the $F(4)$ oscillator are necessarily singular (but square normalizable) at the origin. Therefore, they require the  \cite{{mt},{ftf}} framework of square integrable functions on the real line (which extends the \cite{cal} and \cite{dff}
quantization prescribing wavefunctions defined 
on the $x\geq 0$ half-line and satisfying the Dirichlet boundary condition at the origin). All component wave functions of the $G(3)$ oscillator are regular on the real line, including the origin.

\section{Conclusions}

Unlike its $F(4)$ counterpart, the $G(3)$ superconformal quantum mechanics does not arise as a  quantization of a classical action.
The reason is the following. The $F(4)$ model \cite{akt} is a quantization of a world-line sigma model formulated in the classical Lagrangian setting. Indeed, the scale-invariant restriction \cite{di} of the global ${\cal N}=8$-invariant sigma model \cite{kuroto} based on the $(1,8,7)$ supermultiplet gives a classical theory which  possesses $F(4)$ as dynamical symmetry.  On the other hand, the analysis in \cite{khto} proves that $G(3)$ is classically realized as a $D$-module  representation on a long $(1,7,7,1)$ supermultiplet; this supermultiplet presents fields of four different scaling dimensions and, in particular, a fermionic auxiliary field. A simple argument shows that the presence of this fermionic auxiliary field prevents the construction of a non-trivial invariant classical action with no higher-derivative terms.
\par
On a technical note, the construction of the $G(3)$ superconformal quantum mechanics (and its associated deformed oscillator) is only made possible by the preliminary embedding of the $7$-dimensional representation of the $g_2$ subalgebra within the $8\times 8$ matrices spanning the $Cl(0,7)$ Clifford algebra. A detailed 
presentation of this embedding, within the octonionic-covariant framework, was given in Section {\bf 2}.\par
The method of the octonionically-induced representations, discussed here and in \cite{akt}, implies the so-called ``quasi-nonassociativity", that is the restriction obtained  on the moduli space of the coupling constants as a consequence of
the non-associative structure. It gets reflected, e.g.,  in the determination of $V$ in (\ref{Vpot}) from the octonionic structure constants. As a consequence, at these critical values, an emergent exceptional dynamical symmetry appears.\par
This work concludes the construction of superconformal quantum mechanical models with exceptional finite Lie superalgebras as dynamical symmetry. What is next?
There is no reason to limit the application of the method of octonionically-induced representations to finite algebras. In \cite{eng} the so-called ``Non-associative" ${\cal N}=8$ Superconformal algebra as an ${\cal N}=8$ extension of the Virasoro algebra was introduced. The term ``non-associative" here refers to the property that the graded Jacobi identities are not satisfied. This feature allows to overcome the constraints on the existence of non-trivial central extensions for Virasoro algebras, which are only allowed up to ${\cal N}\leq 4$ (see the classifications given in
\cite{kvdl} and \cite{gls}) if the graded Jacobi identities are assumed. The \cite{eng} ${\cal N}=8$ Non-associative Superconformal Algebra is recovered \cite{caroto} via a Sugawara construction of an octonionic ${\cal N}=8$ superaffine algebra of Mal'cev type. It is therefore a natural candidate to explore the consequences of the octonionically-induced representations in an infinite dimensional setting.\par
Probably, the most promising application of octonionically-induced representations is in connection with the octonionic $M$-algebra introduced in \cite{lt}. Based on the \cite{oku} octonionic realizations of  Clifford gamma matrices, it gives surprising features like a $5$-brane sector which is no longer independent from the particle and $2$-brane sectors as in ordinary $M$-algebra. Its bosonic subalgebra consists of $4\times 4$ octonionic-valued Hermitian matrices. This poses problems for the consistency of its quantization because this structure does not satisfy
the Jordan algebra's axioms. It is worth recalling that the only genuine non-associative system which is quantized within the Jordan's framework is the algebra of $3\times 3$ Hermitian octonionic matrices introduced in \cite{jnw}.  A detailed analysis of the quantization of this model was given in \cite{gpr}. Interestingly, at the end of decade of 1960s, P. Jordan himself investigated the possibility of maintaining a consistent quantization even when Jordan's axioms are relaxed. In this context, the example of the $4\times 4$ Hermitian octonionic matrices was investigated as a toy model (see \cite{lrh} for an historical account of this attempt).
The method of octonionically-induced representations can offer an alternative approach to derive a consistent quantization of the $4\times4$ Hermitian octonionic matrices and of the octonionic $M$-algebra.
It is a completely uncharted territory which deserves unravelling.
 {~}~
\\~
\\~\\~\\~
\par {\Large{\bf Acknowledgments}}
{}~\par{}~\par
I am grateful to Z. Kuznetsova for discussions.
This research was supported by CNPq (PQ grant 308095/2017-0).

\end{document}